\documentclass{ws-mpla}

\begin{document}

\markboth{M. C. B. Abdalla, A. A. Bytsenko and M. E. X.
Guimar\~aes} {Formation of Wakes by Chiral Conducting Cosmic
Strings}
%
\catchline{}{}{}{}{}
%

\title{FORMATION OF WAKES BY CHIRAL CONDUCTING COSMIC STRINGS}

\author{M. C. B. ABDALLA}

\address{Instituto de F\'{\i}sica Te\'orica/UNESP,
S\~ao Paulo, SP, Brazil\\
mabdalla@ift.unesp.br}

\author{A. A. BYTSENKO}

\address{Departamento de F\'{\i}sica, Universidade Estadual de Londrina, Paran\'a,
Brazil\\
abyts@uel.br}

\author{M. E. X. GUIMAR\~AES}
\address{Departamento de Matem\'atica, Universidade de
Bras\'{\i}lia, Bras\'{\i}lia, Brazil \\
marg@unb.br}

\maketitle

\pub{Received (Day Month Year)}{Revised (Day Month Year)}

\begin{abstract}

Chiral cosmic strings are naturally produced at the end of D-term
inflation and they present very interesting cosmological
consequences. In this work, we investigate the formation and
evolution of wakes by a chiral string. We show that, for cold dark
matter, the mechanism of forming wakes by a chiral string is
similar to the mechanism by an ordinary string.
\keywords{Topological defects; formation of large scale
structure.}

\end{abstract}
\ccode{PACS Nos.: 98.80.Cq, 98.70.Vc}

\section{Introduction}

Topological defects are predicted in many gauge models as
solitonic solutions resulting from spontaneous breaking of gauge
or global symmetries. Among all these solutions, cosmic strings,
and in particular those with conducting properties, have a number
of cosmological applications.\cite{vil} The possible scenarios for
structure formation by cosmic strings depend very much in the
detail of the model. One relevant mechanism to understand the
structure formation by these configurations involves long strings
moving with relativistic speed in the normal plane, giving rise to
velocity perturbations in their wake.\cite{silk} The mechanism of
forming wakes has been considered by many authors in both Einstein
and Scalar-Tensor theories of gravity in the neutral string case
\cite{vacha1,vacha2,peri,guima1} and in the conducting string case
.\cite{puy,andre1,andre2}

Recently, it has been shown that for strings like those proposed
by Witten \cite{witten} (i.e, current-carrying strings) there
appear logarithmic terms  and the mechanism of forming wakes can
breakdown.\cite{andre1,andre2} Inclusion of a current-like effect
modifies tremendously the internal structure of a cosmic string in
such a way that new states can be reached. This is due to the
breaking of the Lorentz boost invariance along the worldsheet
allowing rotating equilibrium configurations called vortons. If
these vortons are stable they can overclose the Universe.
\cite{carter,peter} The new feature pointed out in the Refs.
\cite{andre1,andre2} is that a long current-carrying string can
also lead to a catastrophe for the underlying theories that
predict them.

Chiral strings arise when Witten's fermionic
zero mode mechanism gives rise to a purely left (or a purely right) moving
modes with a null current. As a step forward in our previous investigation,
and as a main goal of this paper, we study the impact of the chirality
property of the string in the mechanism of
wakes. For this purpose, we will use
what can be called the ``canonical cold dark matter model''. That is,
we will mimic the mechanism of wakes based on linear adiabatic
perturbations (the Zel'dovich approximation) and cold dark matter flowing past
the chiral string. This letter is outlined as follows.
In the section 2, we briefly recall some general properties of a chiral
string, giving particular attention to its metric. In the section 3, we
treat the mechanism of forming wakes by a chiral string by means of the
Zel'dovich approximation. We finally end up our paper with some conclusions
and future perspectives in the section 4.

\section{Some General Properties of a Chiral Cosmic String and its Metric}


In this section, we will briefly review some properties and the
metric around a chiral cosmic string. We will mainly follow the
results presented in the paper.\cite{steer1} The stress energy
tensor for the infinite string is given by:
\begin{eqnarray}
\label{energy}
T^{\mu\nu} & = & m^2 diag (1,0,0,-k)\delta(x)\delta(y), \nonumber \\
T^{03} & = & m^2 \frac{1-k}{2}\delta(x)\delta(y)
\end{eqnarray}
$k$ caracterizes the state of the string: $k=0$ corresponds to the
maximal charged strings and $k=1$ corresponds to the ordinary, neutral string.
It can happen that $k$ is not constant, but we will not deal with this
case here. The off-diagonal terms are presented because of the null current of the
string and they represent their momentum.
 If $k=1$ we reduce our case to the case of  an ordinary cosmic string,
in which $T^{0}_{0}=T^{3}_{3}$ and the off-diagonal terms vanish.

In the weak-field approximation, $g_{\mu\nu}= \eta_{\mu\nu} +
h_{\mu\nu}$ where $\mid h \mid \ll 1$, and in the de Donder gauge,
\cite{donder} we have:
\begin{equation}
\label{h}
\Box h_{\mu\nu} = 16\pi G (T_{\mu\nu} -\frac{1}{2}\eta_{\mu\nu} T)
\end{equation}
Writing these equations in the cartesian coordinates system in
which $r^2 = x^2 + y^2$, the solution for Eqs. (\ref{h}) are:
\begin{eqnarray}
h_{tt}& = & -h_{tz}  =  h_{zz} \equiv X(r,k) = 4Gm^2 (1-k)\ln (r/r_0)
 \nonumber \\
h_{xx} & = & h_{yy} \equiv Q(r,k) = 4Gm^2 (1+ k) \ln (r/r_0)
\end{eqnarray}
where $r_0$ is an integration constant which can be interpreted as
the width of the string. The metric of a chiral cosmic string can
be simplified by means of the coordinate
transformation:\cite{vilenkin2}
\begin{equation}
(1 - Q(r,k))r^2 = (1 - 2Gm^2(1+k))^2 R^2
\end{equation}
which gives the final form for the metric as:
\begin{eqnarray}
\label{metric}
ds^2 & = & dt^2 (1+ X(R,k)) - dz^2 (1-X(R,k))
- dR^2 \nonumber \\
& & - (1-2Gm^2(1+k))^2R^2d\theta^2  - 2X(R,k)dtdz \, .
\end{eqnarray}
Just as in the case of ordinary string, this metric also has a deficit angle
given by
\begin{equation}
\delta(k) = 4\pi G m^2 (1+k) \, ,
\end{equation}
which is now $k$-dependent.

With the metric (\ref{metric}) we can easily calculate the gravitational
force on moving test particles
\begin{equation}
\label{force}
F(R,k) = -\frac{4G}{R}\frac{1-k}{1+k} \,\, ,
\end{equation}
and the velocity perturbation
\begin{equation}
\label{velocity}
u = 8\pi G m^2 v_s \gamma + \frac{4\pi G}{v_s \gamma}\left ( \frac{1-k}{1+k}
\right) \,\, ,
\end{equation}
where $v_s$ is the string's normal velocity through matter and $\gamma =
(1-v_s^2)^{-1/2}$.
Both the force and the velocity are $k$-dependent and have their maximum for
$k=0$ (the vortons case). Thus, one might expect that chiral
strings wih large charge will be more effective in the formation of wakes.
In the next section we will analyse the accretion problem due to this  string.

\section{Formation and Evolution of the Wakes and the Zel'dovich
Approximation}

In this section we will study the mechanism of formation and evolution of
wakes by a chiral string, with metric (\ref{metric}). As usual and to
simplify our problem, we will consider a situation in which cold dark matter
composed by non-relativistic collisionless particles move past a long string.
In this simple case, we can apply the Zel'dovich approximation which
consists in considering the Newtonian accretion problem in an expanding
Universe using the method of linear perturbations.

We start by considering the velocity perturbation given by
Eq. (\ref{velocity}). In this equation, the first term is equivalent to the
relative velocity of particles flowing past the string. The second term,
appears because of the chirality of the string and it vanishes when $k=1$.

Let us suppose now that the wake was formed at $t_i > t_{eq}$, where
$t_{eq}$ is the time of equal matter and radiation. The
physical trajectory of a dark particle can be written as
\begin{equation}
\label{traj}
h(\vec{x}, t) = a(t) [ \vec{x} + \psi(\vec{x}, t)]
\end{equation}
where $\vec{x}$ is the unperturbed comoving position of the particle
and  $\psi(\vec{x}, t)$ is the comoving displacement developed as a
consequence of the gravitational attraction induced by the wake on the
particle. Suppose, for simplification, that the wake is perpendicular
to the $x$-axis (assuming that $dz=0$ in the metric (\ref{metric})
and $ r = \sqrt{x^2 + y^2}$) in such a way that the only
non-vanishing component of $\psi$ is $\psi_x$. Therefore, the
equation of motion for a dark particle in the Newtonian limit
is
\begin{equation}
\label{newton}
\ddot{h} =  - \nabla_h \Phi
\end{equation}
where the Newtonian potential $\Phi$ satisfies the Poisson equation
\begin{equation}
\label{poisson}
\nabla_h^2 \Phi = 4\pi G_0 \rho
\end{equation}
where $\rho(t)$ is the dark matter density in a cold dark matter universe.
For a flat universe in the matter-dominated era, $a(t) \sim t^{2/3}$.
Therefore, the linearised equation for $\psi_x$ is
\begin{equation}
\label{psi}
\ddot{\psi} + \frac{4}{3t}\dot{\psi} - \frac{2}{3t^2}\psi = 0
\end{equation}
with appropriated initial conditions: $\psi(t_i) = 0$
and $\dot\psi(t_i) = -u_i$. Eq. (\ref{psi}) is the Euler
equation whose solution is easily found
\[
\psi(x,t) = \frac{3}{5}\left[\frac{u_i t_i^2}{t} - u_i t_i
\left(\frac{t}{t_i}\right)^{2/3}\right]
\]
Calculating the comoving coordinate $x(t)$ using the fact that
$\dot{h} = 0$ in the ``turn around"\footnote{The moment when
the dark particle stops expanding with the Hubble flow and starts
to collapse onto the wake.}, we get
\begin{equation}
\label{comoving}
x(t) = - \frac{6}{5} \left[ \frac{u_i t_i^2}{t} - u_i t_i
\left(\frac{t}{t_i}\right)^{2/3}\right]
\end{equation}
With the help of (\ref{comoving}) we can compute both the
thickness $d(t)$ and the surface density $\sigma(t)$ of the
wake.\cite{vil} We have, then, respectively (to first order in
$G$)
\begin{eqnarray}
\label{thick}
d(t) & \approx & \frac{48}{5} \left ( \frac{t}{t_i} \right )^{\frac{1}{3}}
\left ( 2\pi G m^2v_s \gamma + \frac{\pi G}{v_s \gamma} \left (
\frac{1-k}{1+k} \right )\right ) \nonumber \\
\sigma (t) & \approx & \frac{8}{5 t} \left ( \frac{t}{t_i}
\right )^{\frac{1}{3}} \left (  2 m^2v_s \gamma + \frac{1 }
{v_s \gamma} \left (\frac{1-k}{1+k} \right )\right )
\end{eqnarray}
Surprisingly enough, we can easily see that, for $k=0$ (the vorton
case), the results obtained above reduce to those already known
for a wiggly string in General Relativity \cite{vacha1} or an
ordinary string in Scalar-Tensor gravity,\cite{guima1} after some
identifications of the constants in the second term in Eq.
(\ref{thick}). This means that, at least for cold-dark matter, the
details of the model (e.g., whether the string is current-carrying
or not or whether the underlying theory of gravity is purely
tensorial or scalar-tensorial) is not relevant.

\section{Conclusions}

Even tough the recent results of the CMB satellite WMAP \cite{wil} indicate
inflationary models as the source of the large-scale structure in the
Universe, cosmic string models still raise some interest in cosmology
\cite{she}.
In particular,
wakes produced by moving strings can provide an explanation for filamentary
and sheetlike structures observed in the universe. A wake produced by the
string in one Hubble time has the shape of a strip of width $\sim v_s t_i$.
In this work we have treated the accreation problem around a chiral string
using the canonical cold-dark matter model. Although both the force
and the velocity reach their maxima values for $k=0$ (vortons), we came to the
conclusion that the formation of wakes is not sensible to the chirality
property of the string, which is expressed in metric (5) by the mixed term
$2X(R,k) dtdz$. We can infer that, most likely, this is due to the
fact that we restricted ourselves to the cold dark matter case.

Nonbaryonic dark matter is either hot or cold depending on whether
the thermal velocity of the dark particles at the time $t_{eq}$ is
large or neglegible. In a model with hot dark matter (say,
neutrinos) and adiabatic density perturbations, the Zel'dovich
approximation is no longer valid, but requires some
adaptation.\cite{peri} We plan to deal with this problem in a
forthcoming paper.

\section*{Ackowledgments}

The authors are grateful to A. L. N. Oliveira, V. C. Andrade and
D. Steer for fruitful discussions. A.A.B. and M.E.X.G. would like
to thank Funda\c{c}\~ao de Amparo \`a Pesquisa do Estado de S\~ao
Paulo (FAPESP/Brazil) for partial support and the Instituto de
F\'{\i}sica Te\'orica (IFT/UNESP) for kind hospitality. The
authors would like to thank the Conselho Nacional de
Desenvolvimento Cient\'{\i}fico e Tecnol\'ogico (CNPq/Brazil) for
partial support.

\section*{References}{99}

\end{document}